\documentclass[prl,twocolumn]{revtex4}
\usepackage{graphicx}
\renewcommand{\v}[1]{{\bf #1}}
\newcommand{\sign}{{\rm sign}}

\newcommand{\w}{{\omega}}

\def\eqa{\begin{eqnarray}}
\def\eea{\end{eqnarray}}
\newcommand{\eq}{\begin{equation}}
\newcommand{\ee}{\end{equation}}

\newcommand{\<}{\langle}
\renewcommand{\>}{\rangle}
\newcommand{\Tr}{{\rm Tr}}

\newcommand{\ra}{\rightarrow}

\newcommand{\al}{\alpha}

\newcommand{\del}{\delta}
\newcommand{\Del}{\Delta}

\newcommand{\ga}{\gamma}
\newcommand{\Ga}{\Gamma}
\newcommand{\ka}{\kappa}

\newcommand{\si}{\sigma}
\newcommand{\Si}{\Sigma}

\newcommand{\cN}{ {\cal N} }

\begin{document}

\title{Theory of tunnelling into a multi-band superconductor: decoherence and interference}

\author{Da Wang, Yuan Wan and Qiang-Hua Wang}

\email[E-mail address:]{qhwang@nju.edu.cn}

\affiliation{National Laboratory of Solid State Microstructures
and Department of Physics, Nanjing University, Nanjing 210093,
China}

\begin{abstract}
By an exact formulation of tunnelling into a multi-band
superconductor in terms of Green's functions, we demonstrate that
the multi-band feature of the iron-based superconductors can lead
to novel interference between Andreev reflections and decoherence
effect of quasi-particles to Andreev reflections in a tunnelling
junction. These effects depend on the relative sign of the gaps
for s-wave pairing, and the sign of the Fermi velocities also
matters for tunnelling along the nodal direction of d-wave gaps.
Experimental probe of such effects could determine both the
pairing symmetry and the relative sign of the gaps on different
bands.
\end{abstract}

  \maketitle

{\em Introduction:} The discovery of iron-arsenide high-Tc
superconductors\cite{hosono} raises a new tide in the research of
superconductivity.\cite{tide} The immediate theoretical question
to ask is what is the pairing symmetry, for which a
group-theoretical analysis has been a first guide.\cite{sysmetry}
As for the pairing mechanism, a plausible scenario is that the
strong spin fluctuations at wave vectors connecting the hole-like
Fermi pockets near the $\Ga$-point and the electron-like pockets
near the M-pockets centered at the corner of the reduced
Brillouine zone (BZ) can lead to repulsive pair scattering
interactions between the two sets of pockets, and therefore
trigger a minus relative sign between gaps on the two sets of
pockets.\cite{spinfluc} The relative sign between the gaps is
therefore a crucial factor to verify/falsify such scenarios. As
for intra-pocket pairing symmetry, the situation is in hot debate.
Angle-resoled photo-emission (ARPES) measurements find a more or
less isotropic gap on each pocket, lending support to s-wave
pairing.\cite{arpes} Nuclear magnetic resonance (NMR)\cite{nmr}
and penetration depth measurements\cite{penetrate} see indications
of nodes in the gap function, implying d-wave pairing. This is
also supported by some specific heat measurements.\cite{specific
heat} However, more recent NMR and penetration depth experiments
seem to favor node-less pairing.\cite{NMRnodeless} In any case the
relative sign is unknown. Tunnelling experiments provide
signatures of both s-wave and d-wave pairing.\cite{tunnel} This
may be attributed to the sample quality, but we shall demonstrate
that the interpretation of the tunnelling data requires extra care
due to the underlying multi-band feature. The crucial idea is that
tunnelling into a multi-band superconductor causes interference
between the bands, an analogue of the well-known effect in
multi-lead mesoscopic systems. Such an effect may encode the
relative sign between the gaps, but is absent in, or only
partially captured by existing theories of tunnelling.\cite{other
theory} Given the complex experimental situation, a theory that
captures the multi-band effect {\it exactly} in tunnelling is
emergently called for.

In the following we first sketch an exact formulation of
tunnelling into a multi-band superconductor in terms of boundary
Green's functions on the leads, which we derive for practical
cases. We then discuss the tunnelling behavior in junctions
involving iron-based superconductors with all of the four Fermi
pockets. We demonstrate that the multi-bands lead to novel
destructive interference between Andreev reflections and
decoherence effect of quasi-particles to Andreev reflections (both
involving different bands). These effects depend on the relative
sign of the gaps for s-wave pairing, and the sign of the Fermi
velocities also matters for tunnelling along the nodal direction
of d-wave gaps. Experimental probe of such effects could therefore
determine both the pairing symmetry and the relative sign of the
gaps on different bands.

\begin{figure}
\includegraphics[width=0.5\textwidth]{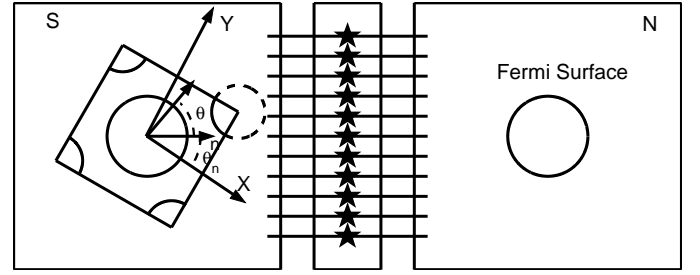}
\caption{Schematic plot of the junction involving a
superconducting L-lead, a scattering region and a normal R-lead.
Here $n$ is the tunnelling direction, $x$ and $y$ are the basis
vectors in the BZ, and $\theta$ is the angle between a Fermi
vector and $n$. The four pieces of the M-pocket at the BZ corners
can be recombined as a closed circle centered at one of the
corners.}
\end{figure}

{\em Exact formulation}: Suppose we are dealing with an N-S
junction with a superconductor carrying multi-bands on the
left-hand lead (L), a normal metal on the right-hand lead (R), and
a scattering center that couples to the $\al$-band of the leads
with amplitudes $t_\al$. The junction geometry is illustrated in
Fig.1. In our case $\al=\Ga_{1,2},M_{1,2}$ on the L-lead and
$\al=N$ on the R-lead. The tunnelling current at bias voltage $V$
can be written as,\cite{tunnel theory} \eqa && I  =\frac{e}{h}\int
d\w\Tr
{G_c^r\Ga^L G_c^a\Ga^R\si_3[f(\w\si_0)-f(\w\si_0-eV\si_3)]}\nonumber\\
&&+\frac{2e}{h}\int d\w G_{c,12}^r\Ga^R_{22}
G_{c,21}^a\Ga^R_{11}[f(\w+eV)-f(\w-eV)],\eea where $G_c^{r(a)}$ is
the retarded (advanced) Green's function on the scattering center,
$\Ga^m=-i(\Si^{m,a}-\Si^{m,r})$ with $\Si^m=\sum_\al (t^m_\al)^2
\si_3 G^m_\al \si_3$ (for $m=L,R$ and applied separately for the
retarded and advanced cases) as the self energy correction to the
scattering center due to coupling to the leads so that
$G_c^{-1}=z\si_0-\sum_m\Si^m$. Here $\si$'s are Pauli matrices in
the Nambu space, and the doubly-numbered subscripts in Eq.(1)
refer to the matrix elements. Moreover, $G_\al$ is the Green's
function in the $\al$-band at the open boundary before attaching
the scattering center. The first term in Eq.(1) describes
quasiparticle contribution, while the second is due to Andreev
reflection which is nonzero if $\Si^L$ contains off-diagonal
components. Throughout this work $z=\w\pm i0^+$ for
retarded/advanced Green's functions, which is left implicit for
notational convenience.

Eq.(1) can be reached by intuitive arguments along the line of
Blonder-Tinkham-Kapwijk (BTK).\cite{BTK} The matrix $T=G_c^r\Ga^L
G_c^a\Ga^R$ describes the transmission probability \cite{tunnel
theory} for particles/holes from R- to L-lead. For electron
transmission, the probability flux is $T_{11}v$, with $v$ the
group velocity. Multiplied by the number of modes $dp/2\pi\hbar$
within a momentum interval $dp$, the total flux is
$T_{11}vdp/h=T_{11}d\w/h$ within an energy interval $d\w$. In
equilibrium this must balance the counter flux from L to R. Under
a bias voltage, an electric current arises as $d\w (e/h) T_{11}
[f(\w)-f(\w-eV)]$. Similarly, the hole transmission contributes a
current $d\w (-e/h)T_{22}[f(\w)-f(\w+eV)]$, where $-e$ accounts
for the positive charge. The off-diagonal parts of $T$ describes
incoming electrons/holes from R transmitted as outgoing
holes/electrons on L. This does not contribute to current, and in
fact cancels out due to charge conservation. On the other hand,
one can take the Nambu components as generalized channels, by
which pairing is nothing but channel hybridization. On the R-lead,
such inter-channel transmission is exactly the Andreev reflection.
In the spirit of $T$, the probability reads $A=G_{c,12}
^r\Ga^R_{22}G_{c,21}^a\Ga^R_{11}$. This contributes an electric
current $d\w (2e/h) A f(\w-eV)$, where the factor of $2$ accounts
for the charge $2e$ transported to L during each Andreev
reflection. Similarly, for incoming holes reflected as outgoing
electrons, the electric current is $d\w (-2e/h) A f(\w+eV)$.
Collecting everything and integrating over $\w$ we arrive at
Eq.(1).

{\rm We emphasize that at this stage the formulation is exact, and
all possible interference effects in our case are exactly
reflected in the self-energies that combines the contributions
from all bands coherently.} To apply the theory we need the
boundary Green's function, the derivation of which is sketched as
follows.

{\em Boundary Green's functions}: Consider an arbitrary band in a
semi-infinite system with a boundary normal
$\hat{n}=(\cos\theta_n,\sin\theta_n)$ relative to the $x$
direction of the BZ (the iron-arsenide-iron direction), as
illustrated in Fig.1. We assume circular Fermi surface, and near
the Fermi momenta we fix the conserved transverse component
$K_{||}=K_f\sin\theta$, where $\theta$ is the angle between $\v
K_f$ and $\hat{n}$. For the M-pockets, $Q_{||}$, the transverse
projection of $\v Q=(\pi,\pi)$, should be added. We then map the
longitudinal degrees of freedom (near the Fermi angles $\theta$
and $\pi-\theta$) to a semi-infinite one-dimensional system
described by the Andreev equations,\cite{BTK} demanding that the
wave function vanishes a unit distance (the lattice constant, for
example) beyond the boundary. These eigen functions and energies
are then used to construct the Green's function in a standard
manner. Another approach is the T-matrix theory for an infinite
plane with an infinite barrier along a line.\cite{tmatrix} We get
the same results using both approaches under the quasi-classical
approximation.

For the s-wave case, the result is independent of $\hat{n}$:
$G(\theta)\sim -\cN J(\theta)\Psi(\Del)$ where $\cN=K_f/2\pi
|v_f|$ is the normal state density of states (DOS) with $v_f$ the
Fermi velocity,
$J(\theta)=2\pi\sin^2(K_f\cos\theta)/|K_f\cos\theta|$ is a
Friedel-like factor due to the open boundary, and
$\Psi(\Del)=(z\si_0+\Del\si_1)/\sqrt{\Del^2-z^2}$, in which
$\sqrt{\Del^2-z^2}=\sqrt{\Del^2-\w^2}\theta(\Del^2-\w^2)\mp
i\sign(\w)\sqrt{\w^2-\Del^2}\theta(\w^2-\Del^2)$ (for
retarded/advanced cases) in practice. Here $\theta(x)$ is the step
function.

For the d-wave case, we consider only two practical cases: If
$\hat{n}$ is along the anti-nodal direction, $\Del_\theta=\Del\cos
2\theta$, and we find $ G(\theta)\sim -\cN
J(\theta)\Psi(\Del_\theta)$ similar to the s-wave case, except for
the $\theta$-dependence in the pairing function.

In contrast, if $\hat{n}$ is along the nodal direction,
$\Del_\theta=\Del\sin 2\theta$, and we find $G(\theta)\sim \cN
J(\theta)\Phi(\Del_\theta)$ with
$\Phi(\Del)=(\sqrt{\Del^2-z^2}\si_0-\Del\sign(v_f)\si_2)/z$, where
$\sign(v_f)=\pm 1$ if $v_f$ is outward (inward) to the center of
the Fermi pocket. The pole at $\w=0$ is a particle-hole mixing
bound state due to the nodal boundary.\cite{boundstate} Notice
that the off-diagonal component in $\Phi$ depends on both the sign
of the gap and the electron/hole nature of the Fermi pocket,
leading to interesting consequence for interference in the Andreev
reflection to be discussed. Finally the Green's function in the
normal lead can be obtained by simply set $\Del=0$.

The above results are applied separately for $G$ in each band,
which is then substituted into Eq.(1) for the conductance
calculation.

{\em Application to junctions with iron-based superconductors}: We
define $\gamma_\al=\cN_\al t_\al^2$ for the four bands
$\al=\Ga_{1,2},M_{1,2}$ on the L-side and for $\al=N$ on the
normal R-side, with $K_\al$ the Fermi radius of the $\al$-pocket.
To simplify the discussion we assume $K_\al=K_f=\pi/2$ for all
bands on the L-lead, and set $\ga_N=\ga$ and
$\ga_{\al=\Ga_i,M_i}=\ka\ga/4$ for a range of values of $\ka$
($\ka=1/0$ corresponds to the transparent/tunnelling limit). We
also set $\ga$ as the unit of energy. For the gap amplitudes on
the four pockets we set $|\Del_{\Ga_1,M_1}|=\Del_1$ and
$|\Del_{\Ga_2,M_2}|=\Del_2$, with $\Del_1=\Del_{\max}$ and
$\Del_2=0.6\Del_{\max}$ in line with existing experiments. We
assume that the gaps on the $\Ga$-pockets are in-phase, so are
those on the $M$-pockets, but the relative sign $\nu=\pm 1$
between the $\Ga$- and $M$-pockets is left free for discussion.

We consider wide junctions, for which the conductance
$\sigma_\theta(\w)=dI_\theta/dV$ is calculated by Eq.(1) (in the
zero temperature limit) for each conserved transverse momentum
$K_{||}=K_f\sin\theta$ (in which $Q_{||}$ should be added for the
M-pockets). The channels unmatched by $K_{||}$ should be shutoff.
The total conductance is then given by
$\sigma(\w)=\int_{-\pi/2}^{\pi/2} \sigma_\theta(\w) K_f\cos\theta
d\theta/2\pi$ per unit area.

We first consider tunnelling along $\hat{n}=(1,1)/\sqrt{2}$. Here
$Q_{||}=0$, and $K_{||}$ can be matched to all Fermi surfaces
under our assumption, because of which we expect novel effects.
For the s-wave case, by using $G_\al$ derived above and under the
given assumptions, we get $\Si^{L}(\theta)=-\ka\ga
J(\theta)\sum_{i=1,2}(
z\si_0-\Del_i\si_1\del_{\nu,1})/2\sqrt{\Del_i^2-z^2}$. a) In the
in-phase case $\nu=1$, the $\si_1$-component in $\Si^L$ is left
intact, and we expect Andreev contribution. However,
quasiparticles at $|\w|>\Del_2$ acts as dissipative source for
(and therefore reduce) the Andreev reflections associated with the
larger gap. This is a {\em decoherence effect}, and is seen as the
dips in Fig.2(a) at energies in between the two gap edges. b) In
the anti-phase case $\nu=-1$, the off-diagonal component cancels
out in $\Si^L$ and consequently in $G_c$. According to Eq.(1) the
Andreev contribution disappears. This is exactly the anticipated
{\em destructive interference effect} among Cooper pairs
transported to the L-side during Andreev reflections on the
R-side. Withinin the smaller gap $\Del_2$ there are no
quasi-particles either so that the conductance is completely zero
at any level of transparency. This behavior is shown in Fig.2(b).
For $|\w|>\Del_2$, quasi-particles from $\Ga_2$- and $M_2$-bands
start to contribute. Of course, both effects becomes less
significant in the tunnelling limit $\ka\ra 0$ where
quasi-particles dominate.

For the $d_{xy}$-wave case $(1,1)$ is the antinodal direction.
According to the previous derivation of $G$, the structure of
$\Si^L$ is similar to that in the s-wave case, except that
$\Del_i$ should be replaced by $\Del_i\cos2\theta$. The
decoherence effect for $\nu=1$ in Fig.3(a) is less visible because
of the size variation in the d-wave gap function. However, the
destructive interference effect for $\nu=-1$ in Fig.3(b) is still
obvious at $\ka\sim 1$.

\begin{figure}
\includegraphics[width=0.5\textwidth]{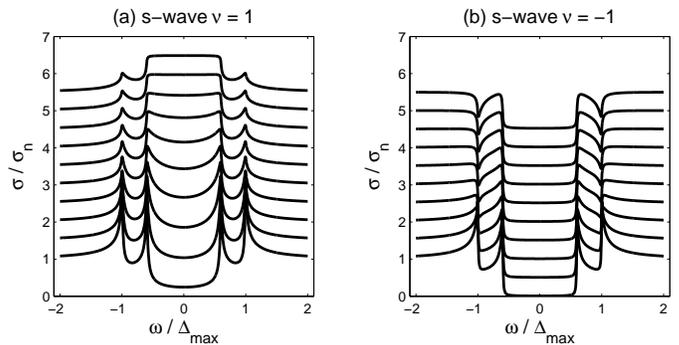}
\caption{Conductance (normalized by the normal state $\si_n$)
versus the bias energy $\w=eV$ for tunnelling along the
$(1,1)$-direction. The relative sign between the s-wave gaps is
$\nu=\pm 1$ in (a) and (b) respectively. In each panel
$\ka=1,0.9,...,0.1$ from top to bottom, and the curves are
vertically shifted for clarity (which approach unity on both sides
of $\w$ otherwise).}
\end{figure}

\begin{figure}
\includegraphics[width=0.5\textwidth]{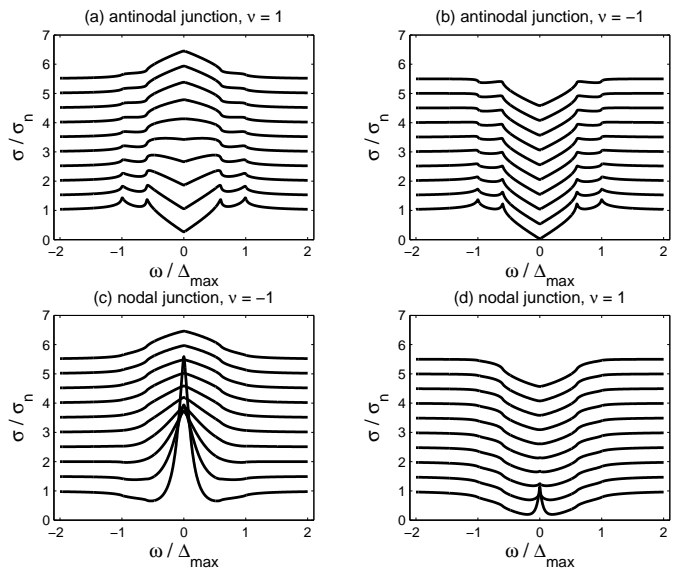}
\caption{The same plot as Fig.2, but for the d-wave case if the
tunnelling direction $(1,1)$ is along the antinodal direction for
(a) and (b), and the nodal direction for (c) and (d).}
\end{figure}

For the $d_{x^2-y^2}$-wave case, $(1,1)$ is the nodal direction.
Here the structure of $\Si^L$ is very different,
$\Si^{L}(\theta)=\ka\ga
J(\theta)\sum_{i=1,2}[\sqrt{\Del_i(\theta)^2-z^2}\si_0+\Del_i(\theta)\si_2\del_{-\nu,1}]/2z$,
where $\Del_i(\theta)=\Del_i\sin(2\theta)$. The extra minus sign
before $\nu$ comes from the fact that the relative sign of the
Fermi velocities is minus for the $\Ga$- and M-pockets. Therefore
we expect destructive interference for $\nu=1$ instead of
$\nu=-1$. a) The result for $\nu=-1$ is presented in Fig.3(c). For
$\ka=1$ the conductance is similar to the corresponding line in
Fig.2(a). With decreaing $\ka$, a central peak shows up and we
checked numerically that the Andreev contribution dominates, due
to the fact that the bound state is a particle-hole mixture. b)
For $\nu=1$, the off-diagonal component cancels out in $\Si^L$,
and the results are shown in Fig.3(d). For $\ka=1$ the conductance
is of broad $V$-shape. The normal (i.e., in the absence of
particle-hole mixing off-diagonal component) bound-state does not
show up here because it has zero group velocity. In the tunnelling
limit $\ka\ra 0$, the normal bound state eventually shows up as a
central peak, since in this limit the DOS on the L-lead is
measured. In addition, no coherence peak occur in (c) and (d) even
in the tunnelling limit, due to the fact that no singularity arise
at the gap edge in $\Phi$. The absence of coherence peaks was
possibly observed.\cite{tunnel}

For tunnelling along $\hat{n}=(1,0)$, the two sets of pockets
match that of the normal lead at different momenta, so that the
$\Ga$- and $M$-pockets can be treated independently. Since within
each set of bands the two gaps are in-phase, the effect of
multi-gaps can be traced in the previous discussion.

We add without going into details that in-plane point-contact on
wide boundary is also helpful. Here one simply averages the
Green's function over $\theta$ first to get the on-site Green's
function, which is then substituted into Eq.(1). There is no
momentum matching constrain. We see that the off-diagonal part of
$G$ survives for the s-wave case and the d-wave anti-nodal
tunnelling case, in which both interference and decoherence effect
can occur. In the case of d-wave nodal tunnelling, the
off-diagonal part cancels out. Furthermore, even out-of-plane
tunnelling is useful. Here one simply use the on-site Green's
function in an infinite plane:
$G_\al=-\pi\cN_\al\<[z\si_0+\Del_\al(\theta)\si_1]/\sqrt{\Del_\al(\theta)^2-z^2}\>_\theta$.
For s-wave gaps, the off-diagonal part still survives and
interference/decoherence occurs, while it cancels out by symmetry
for all kinds of d-wave gaps and only quasi-particles contribute
to the conductance.

We remark that in the real system, the sizes of the four Fermi
pockets differ slightly. The coupling between the scattering
center and each band may also differ. These factors may weaken the
effects discussed so far, but should still be discernable.

{\em Conclusions}: We presented an exact formulation of tunnelling
into a multi-band superconductor, and derived the relevant
boundary Green's functions. Applying to iron-based superconductors
we demonstrate that the multi-band feature can lead to novel
destructive interference between Andreev reflections and
decoherence effect of quasi-particles to Andreev reflections (both
involving different bands) in a tunnelling junction. These effects
depend on the relative sign of the gaps for s-wave pairing, and
the sign of the Fermi velocities also matters for tunnelling along
the nodal direction of d-wave gaps. We propose that in-plane wide
junction and point-contact junction, as well as out-of-plane
contact junction can be used to probe such effects, and therefore
provide decisive information on the pairing symmetry and the
relative sign of the gaps on the various bands.

While finalizing this work, we become aware of some recent
interesting theories.\cite{recent}

Discussion with Bai-Geng Wang is highly appreciated. This work was
supported by NSFC 10325416 and 10734120, the Fok Ying Tung
Education Foundation No.91009, the Ministry of Science and
Technology of China (under the Grant No. 2006CB921802 and
2006CB601002) and the 111 Project (under the Grant No. B07026).

\end{document}